*Баяк И.В.*

## О геометрии плазменного реактора

*Резюме:* Представлена идея управляемого реактора, предназначенного для ядерного синтеза, осуществляемого в вершине конуса, образованного вихревым потоком низкотемпературной плазмы.

Обычно для грубой оценки температуры реакции ядерного синтеза используют приблизительное равенство $\dfrac{m_1\langle v\rangle^2}{2}+\dfrac{m_2\langle v\rangle^2}{2}\cong\dfrac{Z_1Z_2e^2}{r}$, где $\langle v\rangle$ - среднеквадратичная тепловая скорость ионов плазмы; $m_1, m_2$ - массы ионов; $Z_1e, Z_2e$ - электрические заряды взаимодействующих частиц, $r\cong 10^{-11}см$, откуда следует, что $\text{T}\cong 10^8\text{K}$. При этом понятно, что приведенная формула не является выражением закона сохранения энергии в чистом виде, т.к. ядерное взаимодействие задается феноменологической величиной $r$, а гравитационная компонента энергии просто не включается в равенство ввиду ее малости относительно кулоновской составляющей. С другой стороны, согласно принципу эквивалентности Эйнштейна истинное тяготение эквивалентно эффективному инерционному тяготению, величина которого предопределяется условиями эксперимента, поэтому тепловая кинетическая энергия должна быть дополнена составляющей эффективной гравитации. Тогда с учетом данного фактора целесообразно рассматривать более гибкое (в смысле вариабельности условий эксперимента) выражение $m^2\langle v\rangle^2+m^2v_0^2\left(\dfrac{R_1}{R_2}\right)^2\cong\dfrac{e^2}{r}$, где $m_1=m_2=m$; $Z_1=Z_2=1$; $v_0$ - абсолютная величина начальной линейной скорости вращения; $R_1, R_2$ - соответственно начальный и конечный радиус вращения ядра. При этом надо понимать, что в энергетический баланс включена псевдо-гравитационная компонента, иначе говоря, здесь учитывается не взаимодействие, вызванное пространственным сближением ядер, а искривление самого пространства, т.е. возмущение вакуума вблизи ядра.

Таким образом, существует хороший резерв для понижения температуры термоядерной реакции, которым мы не преминем воспользоваться при конструировании игрушечного реактора, т.е. реактора, находящегося в первом приближении к действующему. В соответствии с данной концепцией предлагается схема, принцип работы которой заключается в формировании вращающегося плазменного потока с геометрией конусообразного тора, активизирующего самоподдерживающуюся термоядерную реакцию, т.е. имеет место последовательность процессов:

1) зажигание холодного плазменно-токового шнура пусковым трансформатором,
2) закрутка потока плазмы радиальным ускорителем,
3) инициализация реакции ядерного синтеза, локализованной в области магнитной пробки и сопровождающейся нагреванием плазменного потока в указанной области в результате интенсификации ее протекания,
4) включение термического тока, вызванного градиентом функции температуры плазмы, определенной на оси плазменного потока по всему периметру рабочей камеры,
5) отвод тепла из области локализации более холодной плазмы в нагрузку потребителя.

## Принципиальная схема реактора

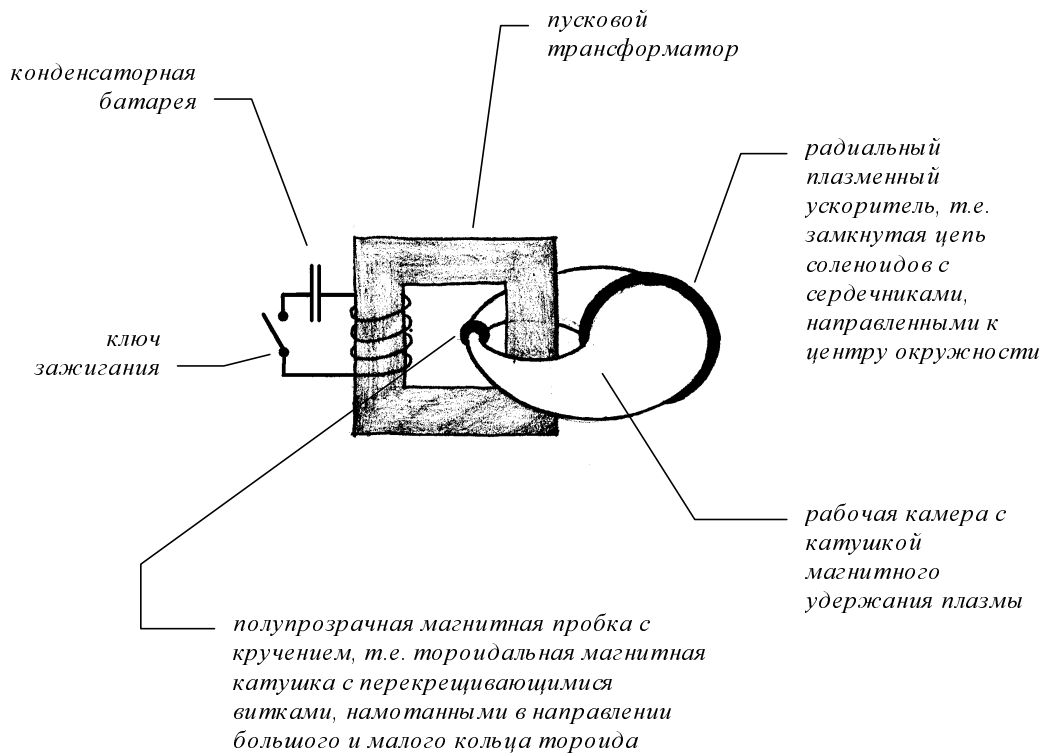

*пусковой трансформатор*

*конденсаторная батарея*

*радиальный плазменный ускоритель, т.е. замкнутая цепь соленоидов с сердечниками, направленными к центру окружности*

*ключ зажигания*

*рабочая камера с катушкой магнитного удержания плазмы*

*полупрозрачная магнитная пробка с кручением, т.е. тороидальная магнитная катушка с перекрещивающимися витками, намотанными в направлении большого и малого кольца тороида*

Несмотря на внешнюю схожесть модели с ТОКАМАКом, предлагаемая концепция имеет свои отличительные особенности, а именно:
- вероятность реакции синтеза ядер возрастает не только за счет повышения температуры и плотности плазмы, но и за счет изменения момента инерции ядра
- устойчивость плазмы соответствует природной стабильности вихревых потоков
- не существует острой проблемы удержания плазмы, поскольку здесь речь идет об удержании в основном холодной плазмы
- реакция синтеза локализована
- поддержка реакции обеспечена автономными процессами
- управление процессом реакции имеет дополнительные составляющие

Итак, хотя на первый взгляд, проектируемый реактор как будто бы должен работать (возможно, даже "наблюдался в работе" в виде произвольно зародившейся шаровой молнии), необходимо все же выполнить некоторые доступные к настоящему моменту ориентировочные расчеты, позволяющие выбрать оптимальную конструкцию реактора. Вместе с тем, отсутствие развитой теории плазмы не должно останавливать нас перед опытной реализацией проекта.